\documentclass[11pt,twoside]{article}

\usepackage{asp2014}

\aspSuppressVolSlug
\resetcounters

\begin{document}

\title{ADARI: Visualizing the quality of VLT data}

\author{C. E. García-Dabó,$^1$ P. Beirao,$^2$ Z. Kostrzewa,$^3$ A. Gabasch,$^1$ B. Wolff,$^1$ M. White,$^5$ M.Deshpande,$^4$ B. Miszalski,$^4$ P. Corcho-Caballero,$^4$ C. Onken,$^5$ A. Heng,$^4$ and M.Gonzalez$^4$}
\affil{$^1$European Southern Observatory – Karl-Schwarzschild-Str. 2 D-85748 Garching bei München, Germany; \email{cgarcia@eso.org}}
\affil{$^2$ATG Europe B.V. - Huygensstraat 34, 2201 DK Noordwijk, Netherlands}
\affil{$^3$Kapteyn Astronomical Institute, University of  Groningen, Landleven 12, 9747 AD Groningen, Netherlands}
\affil{$^4$Australian Astronomical Optics (AAO), Macquarie University, 105 Delhi Rd, North Ryde NSW 2113, Australia}
\affil{$^5$Advanced Instrumentation and Technology Centre, Research School of Astronomy and Astrophysics, The Australian National University, Mount Stromlo Observatory, Canberra ACT 2611, Australia}

\paperauthor{García-Dabó}{cgarcia@eso.org}{0000-0001-5022-3024}{European Southern Observatory}{}{Garching bei München}{}{D-85748}{Germany}
\paperauthor{Gabasch}{}{}{European Southern Observatory}{}{Garching bei München}{}{D-85748}{Germany}
\paperauthor{Wolff}{}{}{European Southern Observatory}{}{Garching bei München}{}{D-85748}{Germany}
\paperauthor{Beirao}{}{}{ATG Europe B. V.}{}{Noordwijk}{}{}{Netherlands}
\paperauthor{Kostrzewa}{}{}{Kapteyn Astronomical Institute,}{}{Groningen}{}{}{Netherlands}
\paperauthor{White}{}{}{Advanced Instrumentation and Technology Centre, The Australian National University}{}{Canberra}{}{}{Australia}
\paperauthor{Deshpande}{}{}{Australian Astronomical Optics (AAO), Macquarie University}{}{North Ryde}{}{}{Australia}
\paperauthor{Miszalski}{}{}{Australian Astronomical Optics (AAO), Macquarie University}{}{North Ryde}{}{}{Australia}
\paperauthor{Onken}{}{}{Australian Astronomical Optics (AAO), Macquarie University}{}{North Ryde}{}{}{Australia}
\paperauthor{Heng}{}{}{Australian Astronomical Optics (AAO), Macquarie University}{}{North Ryde}{}{}{Australia}
\paperauthor{Gonzalez}{}{}{Australian Astronomical Optics (AAO), Macquarie University}{}{North Ryde}{}{}{Australia}
\paperauthor{Corcho-Caballero}{}{}{Australian Astronomical Optics (AAO), Macquarie University}{}{North Ryde}{}{}{Australia}



\begin{abstract}
ADARI (Astronomical DAta Reporting Infrastructure) is a system designed for creating graphical reports of astronomical data so that the quality of these products can be assessed. It has been designed from the ground up to be backend-agnostic, meaning the same ADARI code can be sent to a web plotting API, or a code-based plotting API, with no alteration.

Quick data inspection is an important feature in data reduction systems. The use cases range from quality control at the telescope, advance quality checks prior to delivering data to the scientists as well as data inspection for users running the pipelines at their home institutes. The goal of ADARI is to deliver the same experience and code for data visualization for all the environments, either running automatically in the Paranal Observatory environment or at the PI premises.

ADARI contains a library that can be used to develop the creation of reports as well as a command line tool (genreport) to execute such reports. Most of the time the reports are generated as part of the execution of a data reduction workflow implemented with EDPS, the new ESO system for automatically organising data from ESO instruments and for running the reduction pipelines on them.
\end{abstract}



\section{ADARI Architecture}
The implementation of ADARI is done in modern Python3, using off-the-shelf python plotting technologies like matplotlib and bokeh. The design allows having a plotting backend agnostic implementation for the reports. That way, the same code can be used to create a PNG or PDF hard-copy with matplotlib or an interactive plot embedded in a web application with bokeh.

ADARI consists of several components:

\begin{itemize}
\item {\textbf adari\_core}. This library provides the base-level plotting routines and the core of the backends implementation. It contains separate Plot objects for configuring each specific plot type, like images, histograms or spectra plots for instance. A higher level object are Panels, which contain a set of related plots which gives an overview for a given dataset. Using a Panel object it is possible to construct rich and complex data presentations, consisting of several plots. Additionally, the Panels can contain "links" between plots, in which a change in a property of a plot, like the axis range, affects how other plot is displayed.

\item {\textbf report agent}. This component is responsible for loading the reports and calling the proper backend for rendering.

\item {\textbf genreport}. This is a command line utility that allows easily to run a report on some given data. It finds the specific instrument report, creates the panels to display and generate the artifacts, like PNG files.

\item {\textbf adari\_datalibs} is a library of pre-packaged Panels for astronomical instruments with similar data types. The goal is to minimize the amount of code need to support the reports for a new instrument and have a uniform system that is easier to maintain. The result is reports with the same look and feel for instruments of the same kind, which helps QC operators to easily analyze the reports. 
\end{itemize}

\articlefigure{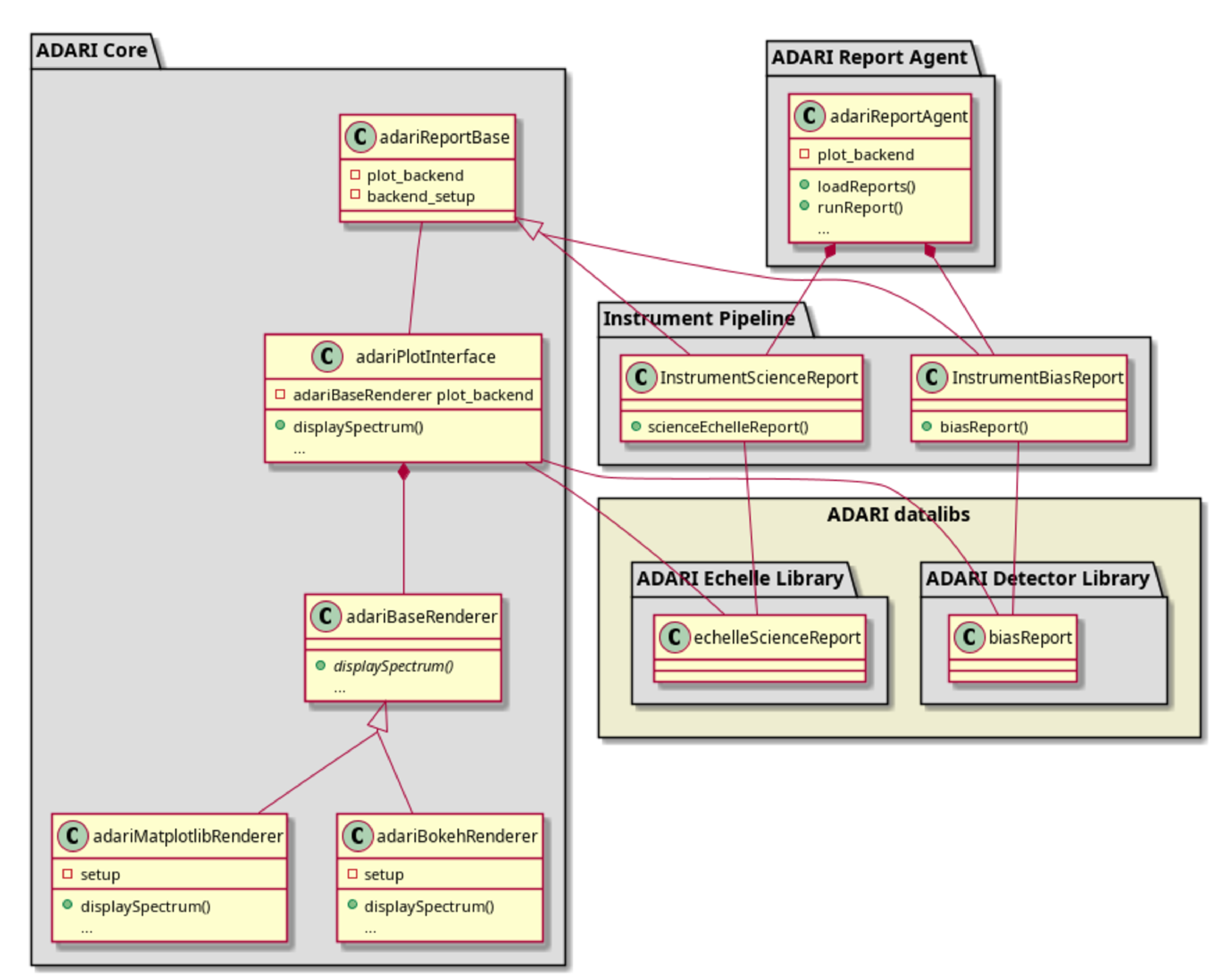}{architecture}{Architecture of ADARI.}

Figure ~\ref{architecture} shows a class diagram of the ADARI components.

\subsection{Panels and reports}
The EDPS \citep{edps} project enables the execution of ESO pipelines with a workflow-based approach. Each of the individual reduction steps, called recipes, usually associate an ADARI report to analyze the data for that particular reduction step, like flat-fielding, wavelength calibration or image stacking. Each report might contain one or several panels, depending on the that that needs to be displayed together in order to asses the quality of the data reduction.

In its current form, the types of data for which ADARI already provides template reports that help in the creation of new reports are the following:

\begin{itemize}
\item {\textbf Echelle data}. There are template reports for echelle flat fields, wavelength calibrations, format checks and order definitions.

\item {\textbf 3D data}. These support IFU flats as well as astrometry calibrations.

\item {\textbf Interferometry}. The panels here include visibilities, phase closure and transfer functions.

\item {\textbf Imaging}. Template reports exist for bias, darks, detector monitoring and flats.

\item {\textbf Raw data}. Additionally, there is a template to display raw data in a consistent way for all instruments.
\end{itemize}

\articlefigure{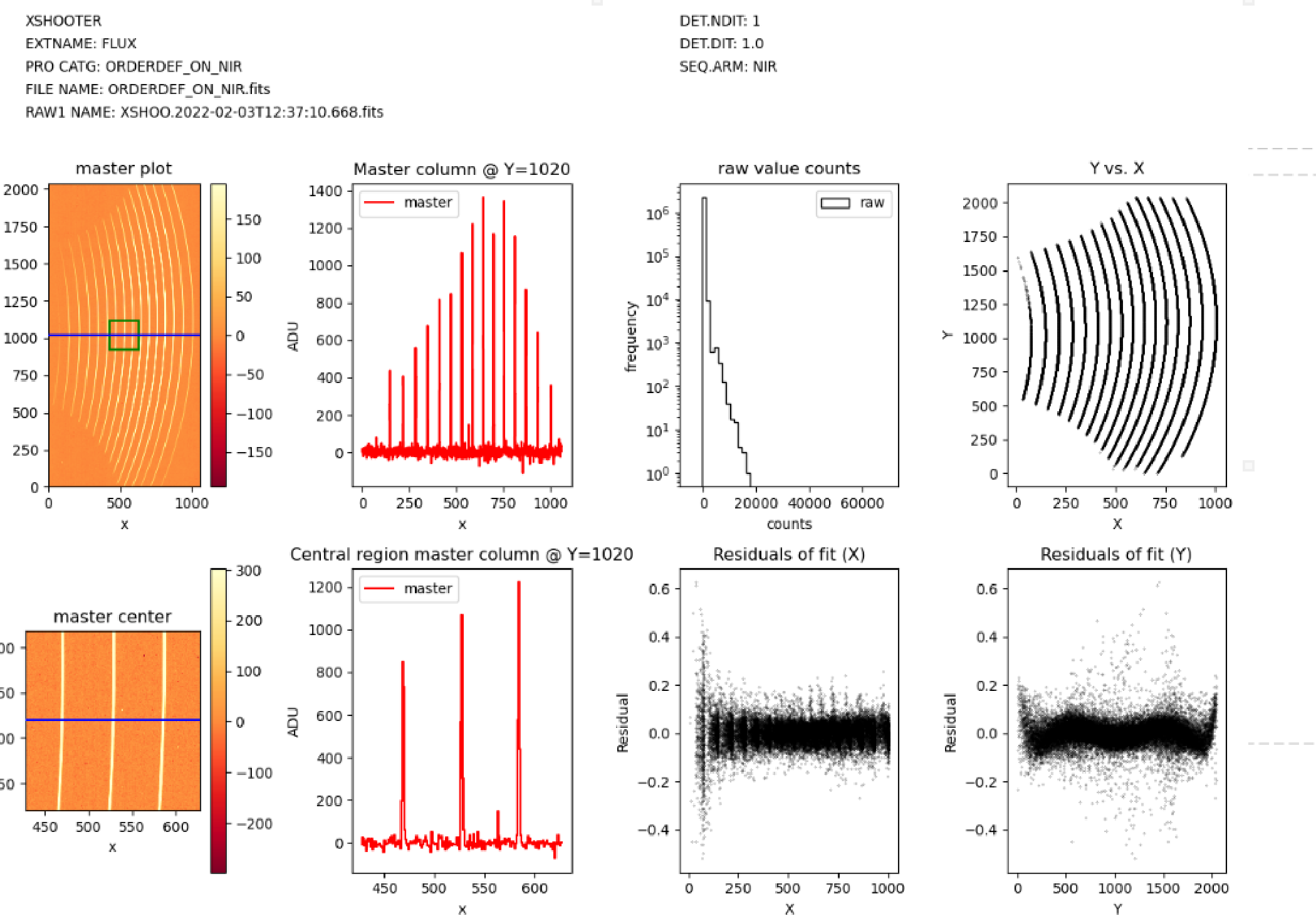}{example}{An example panel created with ADARI.}

Figure \ref{example} shows a wavelength calibration panel for the X-Shooter instrument created with ADARI.

\section{Testing strategy}
The ADARI libraries are equipped with a set of unit tests for each kind of plot objects. These are basic unit tests that ensure that all the plotting functionality is working as expected. Note that the Plot objects are backend-agnostic, so there is nothing there that exercises the real rendering of the plots.

To have an easy comparison of the generated plots, a JSON backend as been implemented. This backend {\sl renders} the plot in a human-readable JSON file that is easy to compare with previous executions. The JSON artifacts are a meta-description of the plot that can be used to easily check differences between the generated plots.

There are over +3000 regression tests that use real data used for production instruments. These tests compare both the PNG as well as the JSON artifacts. They are extremely useful to avoid regression in the plots when a code refactoring happens in ADARI.

\section{Future directions}

ADARI reports are being implemented for all operational VLT instruments and currently 12 out of 16 have been fully deployed in the ESO QC environment. In the course of 2025, additional reports for the ESO internal science products will also be produced.

Most of the effort so far has been focused in producing reports for QC assessment. The development in the coming months will address the creation of interactive plots and its inclusion within the EDPS graphical interface. The target audience for that functionality will be users of ESO data who want to reduce and analyze data using the ESO pipelines in an interactive fashion.

\bibliography{P817}  

\end{document}